# Spontaneous Symmetry Breaking and Strong Deformations in Metal Adsorbed Graphene Sheets

**A. F. Jalbout**

*Departamento de Investigacion en Fisica, Universidad de Sonora, Hermosillo, México*

**Y.P. Ortiz**

*Instituto de Ciencias Físicas, Universidad Nacional Autónoma de México, Cuernavaca, México*

**T.H. Seligman**

*Instituto de Ciencias Físicas, Universidad Nacional Autónoma de México, Cuernavaca, México and Centro Internacional de Ciencias, Cuernavaca, México*

*Abstract*

We study the adsorption of Li to graphene flakes described as aromatic molecules. Surprisingly the out of plane deformation is much stronger for the double adsorption from both sides to the same ring than for a single adsorption, although a symmetric solution seems possible. We thus have an interesting case of spontaneous symmetry breaking. While we cannot rule out a Jahn Teller deformation with certainty, this explanation seems unlikely and other options are discussed. We find a similar behavior for Boron-Nitrogen sheets, and also for other light alkalines.

**Introduction**

Graphene has received a great deal of attention, which has more recently extended to boron nitride sheets (BNS) with a similar structure but reduced symmetry. Nowadays, the principal importance of graphene is no longer centered on being the building block of nanotubes. Indeed graphene and BNS have acquired a significance of their own. Thus shaping the sheets and influencing the charge distribution is of greatest interest. Concerning the second point, adsorption of electron donors such as lithium atoms has been shown to be efficient on small sheets [1] and quasi 1-D molecules such as polyacenes and polyphenyls [2]. Increasing computer power makes explorations on a larger scale possible.

In previous studies we demonstrated that in the $Li^+$ case, a charge of ~0.8 $e$ is transferred to the surface of the fullerene ($C_{60}$) molecule [3] and small graphene sheets [1]. Modification of the surface electron configuration was also shown to yield increased reactivity. Based on these and similar results for nanotubes [4-6], we shall explore adsorption of alkaline atoms, particularly lithium to graphene and BNS.

The basic results of our previous calculations suggests that in fullerenes the $Li@C_{60}$ (whereby Li is encapsulated in the $C_{60}$ frame) system can roughly be viewed as $Li^+@C_{60}^-$ since the valence electron density of the Li atom to a large extent is transferred to the fullerene surface [3]. Analysis of the charge population showed that the majority of this density is concentrated in a distinct region of the surface, which implies that the adsorption occurs a specific point. In the present analysis we shall discuss how charge mediated bond stretching caused by metal adsorption can influence the physical (and chemical) properties of carbon and boron nitride sheets.

Among these properties we shall find, as the central result of this letter, spontaneous symmetry breaking. This effect can be explained in terms of a qualitative discussion of bond stretching while results on quasi 1-D molecules such as polyacenes and polyphenils showing similar distortions may indicate some quantum-phase transition, which is hidden by the fact that adsorption happens in finite steps.

We must recall, that the sheets under discussion have an in plane symmetry of translations, rotations and reflections as well as a mirror symmetry at the plane of the sheet. Minor wrinkles in the unperturbed sheets have been the subject of discussions, and might be due to finite size effects. In our theoretical discussion we disregard them, but they do occur in our numeric, yet on a much smaller scale, then the effects we discuss.

Unsurprisingly our calculations reveal a deformation of the lattice, which violates this mirror symmetry if we adsorb a single atom. If two atoms are adsorbed they can either be fixed to independent sites on the same or opposite side of the sheet, or the second adsorbed atom can find a stable position on the other side of the sheet opposite to the first one. The surprise is, that the deformation in the latter case is much larger than in the former. For one adsorbed atom any Born-Oppenheimer Hamiltonian will violate the mirror symmetry on a scale greater then the known wrinkles, but if two atoms are adsorbed on opposite sides of the sheet a symmetric configuration could exist. This symmetric nuclear configuration is unstable in any of the approximations we used, i.e. in Hartree-Fock calculations with and without some electron correlations as well as in density functional computations. For graphene we shall show this effect for increasing sizes of flakes from 6 to 20 rings, to make sure that we do not see merely see finite size effects. The symmetry breaking is manifest throughout this range.

**Results and Discussion**

All of the quantum chemical computations were performed closing the sheets with hydrogen atoms. The GAUSSIAN03 [8] codes were used. Since the systems are relatively large the geometry optimizations were performed with the Hartree-Fock (HF) method. Higher-level calculations (DFT and electron correlations over the Hartree Fock results) were performed for specific geometries to confirm consistency of the mean field results. These tests were important, because experiments on the systems explored are not yet available.

To do the calculations we have selected the STO-3G* basis which is an acronym for Slater-Type-Orbitals simulated by a superposition of three Gaussians in this work. The coefficients of the Gaussian functions are adjusted to yield reasonable fits to the Slater orbitals. Some of the smaller sheets were calculated with larger basis sets to ensure that the results are reliable. The B3LYP method coupled to the same basis sets was used for the confirming calculations. This method uses the Becke three parameter hybrid functionals which are composed of $A*E_X^{Slater}+(1-A)*E_X^{HF}+ B*\Delta E_X^{Becke}+E_C^{VWN}+C*\Delta E_C^{non-local}$ by which A, B, and C are the constants determined by Becke and suitable for the G1 molecular set. Furthermore we have performed calculations and tested our results with sodium and potassium for which they were consistent with those for lithium. They will not be reported in this manuscript.

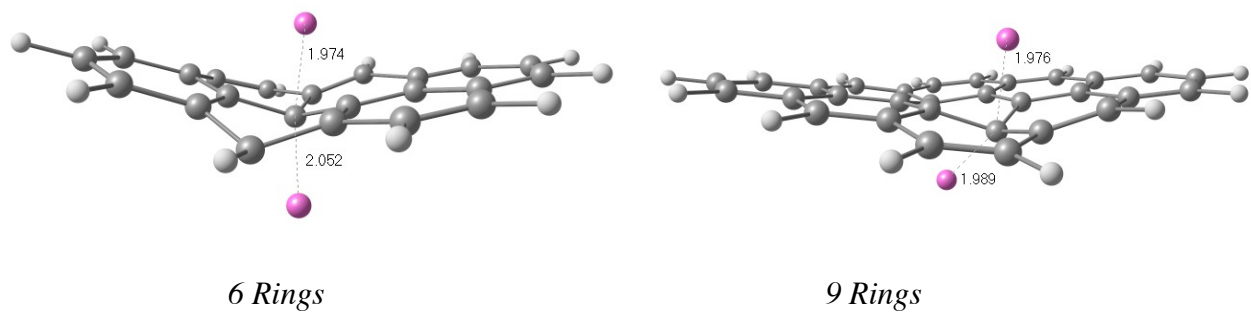

6 Rings                                   9 Rings

**Figure 1.** Double Li atom adsorption to 6,9 sheet system.

The single Li adsorption energy to the 6-ring system is around 59.1 kcal/mol with an intermolecular separation of around 2.4 Å for one of the Li atoms to the surface. At the HF/STO-3G level of theory the Mulliken partial charge on Li is around 0.26$e$. For the shown double adsorption we can observe the strong symmetry breaking in both species. Yet the adsorption energy for the second Li atom is only 11.13 kcal/mol. Despite of this the distances of the Li atoms to the sheet are smaller.

We have performed a test HF/6-311G** natural orbital population analysis (NOPA) which yields a similar partial charge on Li of 0.15$e$ and determines that the transferred electron density is occupying a localized region on the hydrocarbon surface. Another important feature is the local nature of the symmetry breaking. While NOPA charges are more accurate it is our intention to discuss general trends upon metal absorption to an extended planar aromatic molecular surface. Calculations of the Mulliken partial charges suggest that the Li atoms equally donate around 0.74$e$ to the molecular surface independent of the surface employed.

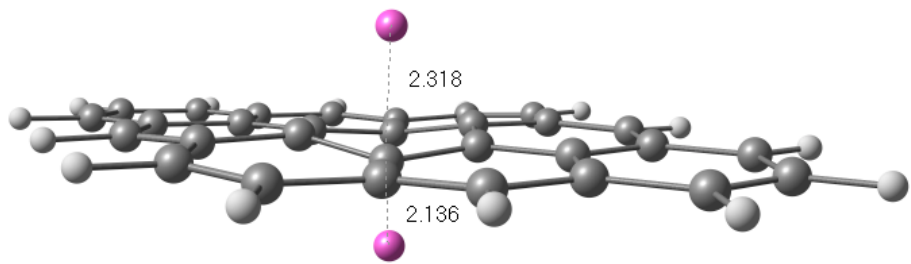

**Figure 2.** The DFT-B3LYP/3-21G* result of double Li adsorption to the nine-ring graphene system.

An increase of the system to 9 rings creates an intermolecular separation of around 2.2 Å. We observe that the Li atoms tend to induce an attractive polarization force when they are near the molecular surface. It is also important to note that the results are consistent within density functional theory (DFT), which is known to overcome some problems of mean field results. The DFT results show similar symmetry breaking as can be seen in Figure 2.

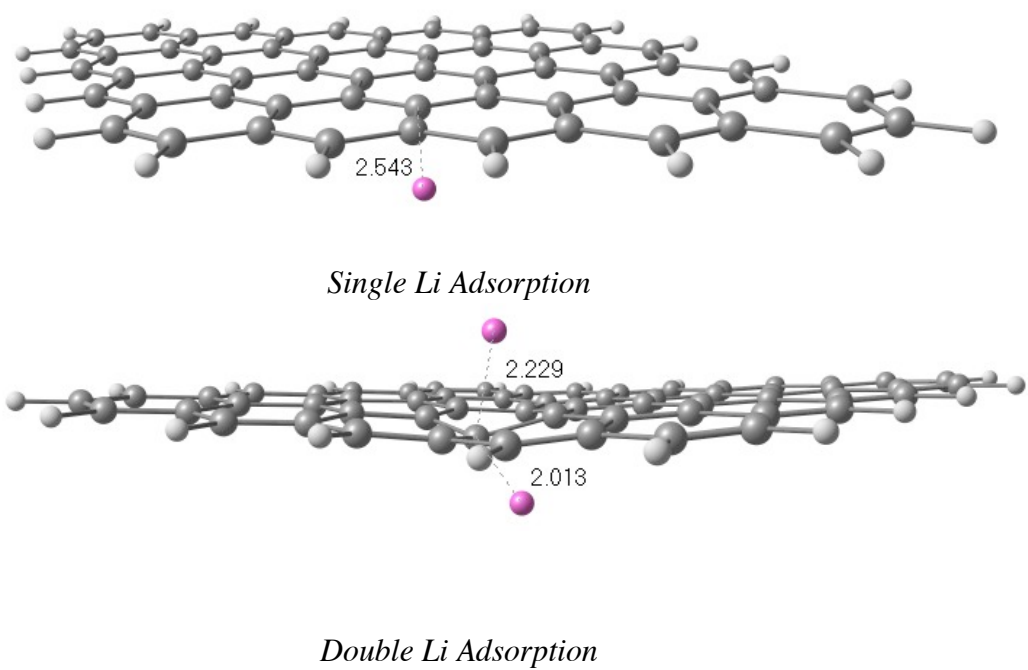

*Single Li Adsorption*

*Double Li Adsorption*

**Figure 3.** Single and double Li adsorption to a 20-ring graphene ring systems

To strengthen the notion that developed above that this is a property of graphene sheets and not of small aromatic molecules we present the adsorption of Li atoms to the 20-ring system. In this case (Figure 3) we show the results of single and double Li adsorption to demonstrate that the effects of single adsorption are much weaker than double adsorption. Note that the distances from the sheet are larger and the deformations are much smaller in the single adsorption case. This DFT result shows how an extended surface exhibits local symmetry distortions. The single Li adsorption case causes comparatively minor distortions to the sheet, but still bigger then any wrinkles far from the adsorption area.

The calculated, deviation between adsorption energies for different sheet sizes is shown in Table 1.

| No. Rings | First Adsorption | Second Adsorption |
|:---:|:---:|:---:|
| 6 | 59.07 | 11.13 |
| 7 | 99.30 | 21.23 |
| 8 | 103.72 | 16.32 |
| 9 | 109.28 | 22.23 |
| 17 | 287.71 | 56.88 |
| 18 | 275.84 | 62.23 |
| 19 | 279.43 | 81.77 |
| 20 | 290.56 | 90.07 |

**Table 1.** Adsorption Energies (kcal mol$^{-1}$) of single and then double Li atoms to the surface of the graphene.

Note that while double adsorption caused strong distortion the main part of the distortion energy is already produced by the adsorption of a single Li atom. Yet clearly the adsorption of the second atom yields additional energy.

A simple understanding of the observed effect is possible. The adsorbed charge lengthens the bonds in the region of localized enhanced electron density. While for a small sheet a deformation in the plane would be possible in a large or infinite sheet this cannot happen and the sheet will have to buckle to either side. The sequence in which the two atoms are adsorbed will obviously determine the side to which the longer bonds will push the site. If additional electronic charge leads to bond stretching, this will be stronger for two, then for one adsorbed atom. As double adsorption to one site from opposite sides is possible this strong symmetry breaking occurs. Recent results [2,8] for quasi1-D molecules made up of benzene-like rings i.e. polyacenes and polyphenils show that such deformation can result without the constraints of embedding in a sheet.

An alternate explanation seems very tempting. It is based on the observation, that a mean field theory of graphene leads to gapless spectrum associated to a massless Dirac equation [9]; this property has been quite well confirmed experimentally [10]. Searching for a deeper understanding, we might suspect, that it is this gapless structure, which is responsible for the instability. To illuminate this point we analyze a boron nitride sheet where, due to the mass-difference, a gap in the mean field spectrum must arise. As can be seen in Figure 4, this sheet displays a very similar deformation, thus excluding such an alternative.

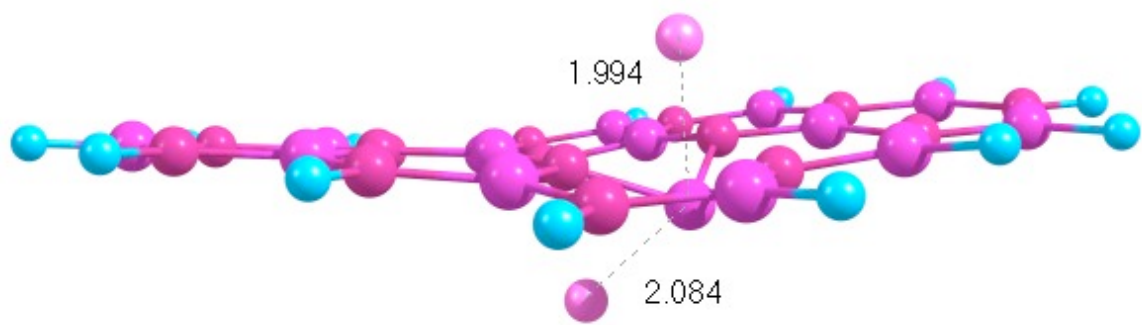

**Figure 4.** Double Li adsorption to a 9 membered BN ring system that clearly shows the spontaneous symmetry breaking.

Indeed Figure 4 presents the results of double Li adsorption to a nine ring BN sheet computed at the DFT-B3LYP level of theory. Interestingly, we see the same results as before with a similar intermolecular separation in the cases calculated. The single Li adsorption energy for the BNS is 15.92 kcal/mol, while the double adsorption energy is 37.3 (DFT) kcal/mol.

Having thus excluded the spectral structure as a source for the instability of the symmetric configuration of the nuclei, we have suspected a cause, similar to the Jahn Teller effect [1]. While it is difficult to exclude ground state degeneracy in a system of such complexity the results on the quasi 1-D molecules indicate that bending of the molecule produces zero derivative of the energy with respect to the bending angle [8]. On the other hand the quasi 1-D results suggest that periodic structures can appear. In Figure 5 we show two double adsorptions to a eight ring flake and we see some regularity emerging, which can be guessed from the behaviour of the 1-D molecules, noting that two polyacene like strips bend in opposite direction. Further caculations along these lines are under way.

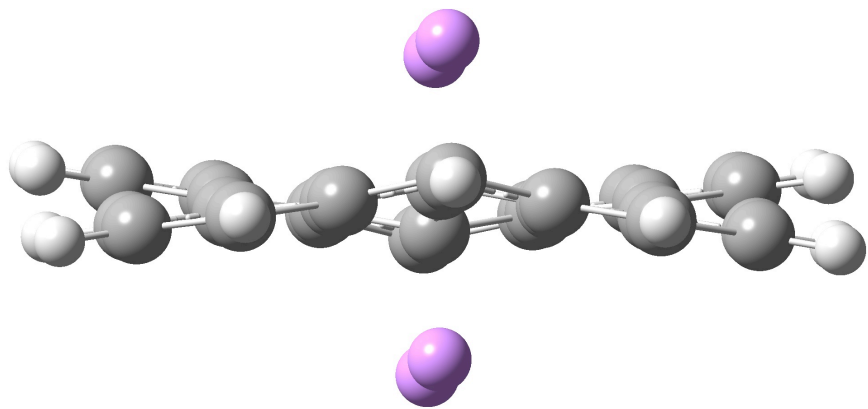

**Figure 5.** Two double Li adsorptions to a 8-ring flake

**Conclusions**

Adsorption of electron donor atoms to graphene and BNS does not only lead to localized charge density modulation in these sheets, but taking advantage of the symmetry breaking mechanism explored above, fairly strong local deformations can be obtained without affecting the general structure of the sheets. Considering, that we can use e.g. an atomic force microscope to deposit electron donors at predetermined points on the sheet, we can design pre-established structures of the charge distribution on this sheet and furthermore we can hope to bend the sheets in desired ways if we have lines of singly or doubly adsorbed alcalines; a first very simple example was shown in Figure 5. This opens an entire new field of mechanical and electrostatic modulation that could be the starting point for many "designer systems", creating a new perspective in mesoscopics. We may even speculate that an appropriate interplay of charge and shape dependence may transfer to the temporal domain and allow the construction of moving nano-devices.

**Acknowledgements**

We appreciate support from CONACyT Project Number 79613 and PAPIIT-UNAM project number IN 114310.

14